\documentclass{anstrans}

\title{Demonstrating Quadratic Monte Carlo Speedup via Quantum Amplitude
       Estimation: Nuclear Engineering Examples}

\author{Jilang Miao$^{*}$, Miaomiao Jin$^{\dagger}$, and Akira Sone$^{\ddagger}$}

\institute{%
$^{*}$$^{\dagger}$Ken and Mary Alice Lindquist Department of Nuclear Engineering,
Pennsylvania State University, University Park, PA 16802 \\
$^{\ddagger}$Department of Physics, University of Massachusetts Boston,
Boston, MA 02125
}

\usepackage{graphicx}
\usepackage{booktabs}
\usepackage{amsmath}
\usepackage{microtype}
\usepackage[section]{placeins}  % \FloatBarrier at each \section automatically
\usepackage{stfloats}            % allows figure* to appear mid-page (top or bottom)

\begin{document}
\maketitle
{\let\thefootnote\relax\footnotetext{%
$^\S$Corresponding author: Jilang Miao, \texttt{jlmiao@psu.edu}}}

%%%%%%%%%%%%%%%%%%%%%%%%%%%%%%%%%%%%%%%%%%%%%%%%%%%%%%%%%%%%%%%%%%%%%
\section{Introduction}

Monte Carlo (MC) methods are the gold standard for high-fidelity reactor
physics calculations, providing accurate predictions of neutron transport,
criticality, and reaction-rate distributions without the geometric or energy
approximations required by deterministic solvers.
Their principal limitation is statistical: the root-mean-square error of a
MC estimator scales as $\sigma_f / \sqrt{N}$, where $N$ is the number of
particle histories and $\sigma_f$ is the standard deviation of the
integrand.
Reducing the error by a factor of ten demands one hundred times more particle
histories, which is a fundamental bottleneck that motivates continued efforts
in variance reduction and parallel hardware~\cite{Martin2012}.

Quantum computing offers a potential solution to this fundamental problem.
Quantum amplitude estimation (QAE), introduced by Brassard
et al.~\cite{Brassard2002} and connected to the general MC speedup
theorem by Montanaro~\cite{Montanaro2015}, achieves an error that
scales as $O(1/T)$, where $T$ is the the number of oracle calls; this represents a quadratic
improvement over the classical $O(1/\sqrt{N})$.
The algorithm has been demonstrated for financial risk
integrals~\cite{Woerner2019,Stamatopoulos2020}, but applications
to nuclear quantities have received limited attention~\cite{Olivier2024}.

This paper demonstrates quantum phase estimation (QPE)-based QAE on a Qiskit~2.2~\cite{Qiskit2024}
statevector simulator for two reactor physics quantities:
(1)~the mean fission neutron yield $\bar{\nu}$, modeled as an expectation
over a discrete yield distribution; and
(2)~the U-238 resonance integral $I_R = \int \sigma_a(E)/E\,dE$,
a physically relevant quantity governing neutron slowing-down in the
epithermal energy range.
Both cases confirm the quadratic convergence advantage and also reveal
practical implementation challenges at the current simulator scale.

%%%%%%%%%%%%%%%%%%%%%%%%%%%%%%%%%%%%%%%%%%%%%%%%%%%%%%%%%%%%%%%%%%%%%
\section{Quantum Amplitude Estimation}

\subsection{Classical MC Baseline}

The target is the expectation $\mu = \mathbb{E}[f(X)] = \sum_x P(x)\,f(x)$,
where $P(x)$ is a probability mass function and $f: X \to [0,1]$ is a
bounded scoring function.
Classical MC draws $N$ independent samples based on $P$ and averages:
$\hat{\mu}_N = N^{-1}\sum_{i=1}^N f(X_i)$.
The mean squared error $\mathrm{MSE} = \sigma_f^2/N$ where
$\sigma_f^2 = \mathrm{Var}[f(X)]$, so halving the error quadruples the
required number of sampling, due to the $1/\sqrt{N}$ scaling.

\subsection{Qubits and Quantum States}

A qubit is a two-level quantum system whose state is a complex superposition
$|\psi\rangle = \alpha|0\rangle + \beta|1\rangle$ with normalization
$|\alpha|^2 + |\beta|^2 = 1$.
Unlike a classical bit, which is definitively 0 or 1, a qubit exists in
both states simultaneously until \emph{measured}, at which point it collapses
to $|0\rangle$ with probability $|\alpha|^2$ or $|1\rangle$ with probability
$|\beta|^2$.
A register of $n$ qubits can represent any superposition over $2^n$ basis
states: $|\Psi\rangle = \sum_{x=0}^{2^n-1} c_x|x\rangle$, with $\sum|c_x|^2 = 1$.
Quantum gates are unitary transformations on this exponentially large state space.
The Hadamard gate $H|0\rangle = (|0\rangle+|1\rangle)/\sqrt{2}$ creates
equal superposition, and rotation gates $R_y(\alpha)|0\rangle = \cos(\alpha/2)|0\rangle +
\sin(\alpha/2)|1\rangle$ allow continuous amplitude control.

\subsection{Encoding the Monte Carlo Problem}

A unitary operator $\mathcal{A}$ can encode both $P(x)$ and $f(x)$ into a
single quantum state by loading $\sqrt{P(x)}$ as amplitudes on the
system register and $\sqrt{f(x)}$ onto an ancilla qubit (denoted with subscript $a$),
\begin{equation}\label{eq:stateprep}
  \mathcal{A}|0\rangle = \sum_x \sqrt{P(x)}\,|x\rangle
  \bigl(\sqrt{1-f(x)}\,|0\rangle_a + \sqrt{f(x)}\,|1\rangle_a\bigr).
\end{equation}
Measuring only the ancilla gives $|1\rangle_a$ with probability
$\sum_x P(x) f(x) = \mu$.
However, if the circuit is only used by repeatedly preparing the state and measuring the ancilla, 
estimating $\mu$ to error $\epsilon$ still requires 
$O(1/\epsilon^2)$ repetitions. 
This gives no asymptotic advantage over classical MC.
The real speedup comes from \emph{avoiding} premature measurement and instead
coherently amplifying $\mu$ before reading it out.

\subsection{Amplitude Amplification and Phase Estimation}

Writing $\mu = \sin^2\!\theta$,
we decompose the prepared state into
a ``good'' component with ancilla state $|1\rangle_a$ and
a ``bad'' component with ancilla state $|0\rangle_a$.
The Grover iterate
$\mathcal{Q} = \mathcal{A}\,S_0\,\mathcal{A}^\dagger\,S_\chi$
acts within the two-dimensional subspace spanned by these two components and 
rotates the quantum state by angle $2\theta$ per application.
Here $S_0 = I - 2|0\rangle\langle 0|$ applies a phase flip to the all-zeros input state
and $S_\chi = I - 2\sum_x|x,1\rangle\langle x,1|$ applies a phase flip to all good states, i.e., all states whose ancilla qubit is $|1\rangle_a$.
Rather than measuring after amplification, which would still give
only a single Bernoulli trial, QPE~\cite{Nielsen2000}
reads off the angle $\theta$ directly from the eigenphase of $\mathcal{Q}$.

QPE uses an $m$-qubit register initialized by a Hadamard on each qubit
(creating the superposition $2^{-m/2}\sum_{c=0}^{2^m-1}|c\rangle$),
followed by controlled-$\mathcal{Q}^{2^k}$ gates conditioned on
register qubit $k$ (for $k = 0, \ldots, m-1$), and finally the
inverse quantum Fourier transform (QFT$^\dagger$).
Intuitively, qubit $k$ controls $2^k$ applications of $\mathcal{Q}$,
accumulating the eigenphase into the register; QFT$^\dagger$ then
concentrates the probability onto the register value that encodes $\theta$.
The total oracle cost is $T = \sum_{k=0}^{m-1} 2^k = 2^m - 1$ applications of $\mathcal{Q}$.

Because $\mathcal{Q}$ has eigenvalues $-e^{\pm 2i\theta}$, the QPE register
peaks at $\hat{y}$ corresponding to eigenphases $\varphi_\pm = \tfrac{1}{2} \pm \tfrac{\theta}{\pi}$.
The shift of $\tfrac{1}{2}$ relative to the conventional QPE ($\pm \tfrac{\theta}{\pi}$) arises from the
$-1$ prefactor; the Qiskit implementation assigns register qubit $k$ to
control $\mathcal{Q}^{2^k}$ (LSB-first), which reverses the conventional
bit ordering~\cite{Nielsen2000} but yields the same probability distribution.
Decoding the most-probable register outcome $\hat{y}$ gives
\begin{equation}\label{eq:decode}
  \hat{\theta} = \pi\!\left|\frac{\hat{y}}{2^m} - \frac{1}{2}\right|,
  \qquad \hat{\mu} = \sin^2\!\hat{\theta}.
\end{equation}
The mode-estimation error satisfies~\cite{Brassard2002}
\begin{equation}\label{eq:bound}
  |\hat{\mu} - \mu| \leq \frac{\pi\,|\!\sin 2\theta|}{T}
                         + \frac{\pi^2}{T^2},
\end{equation}
so the squared error scales as $O(1/T^2)$, a quadratic improvement
over the classical $O(1/N)$.

%%%%%%%%%%%%%%%%%%%%%%%%%%%%%%%%%%%%%%%%%%%%%%%%%%%%%%%%%%%%%%%%%%%%%
\section{Case Study}

\subsection{I. Mean Fission Neutron Yield}

\paragraph{Problem setup}
As a minimal benchmark, we estimate the mean fission neutron yield
$\bar{\nu} = \mathbb{E}[\nu]$ for a simplified four-outcome model:
$\nu \in \{0, 1, 2, 3\}$ neutrons per fission with probability mass
$P = (0.1,\, 0.2,\, 0.3,\, 0.4)$, giving $\bar{\nu} = 2.0$ exactly.
First, we set the function $f(\nu) = \nu/3 \in [0,1]$ to scale the yield into $(0,1)$. Then, $\mu = \mathbb{E}[f(\nu)] = 2/3$ and
$\bar{\nu} = 3\mu$.
The four possible values of $\nu$ are encoded with
$n=\lceil \log_2 4\rceil = 2$ state qubits, using the basis states
$|00\rangle, |01\rangle, |10\rangle,$ and $|11\rangle$ for
$\nu=0,1,2,$ and $3$. One additional ancilla qubit is used to encode
$f(\nu)$ as the probability amplitude of the $|1\rangle_a$ state.
For each $m$, 8192 shots are taken from the QPE circuit and the
most-probable register outcome (mode of $\hat{y}$) is decoded via
Eq.~\eqref{eq:decode}.

\begin{figure*}[t]
  \centering
  \includegraphics[width=0.96\textwidth]{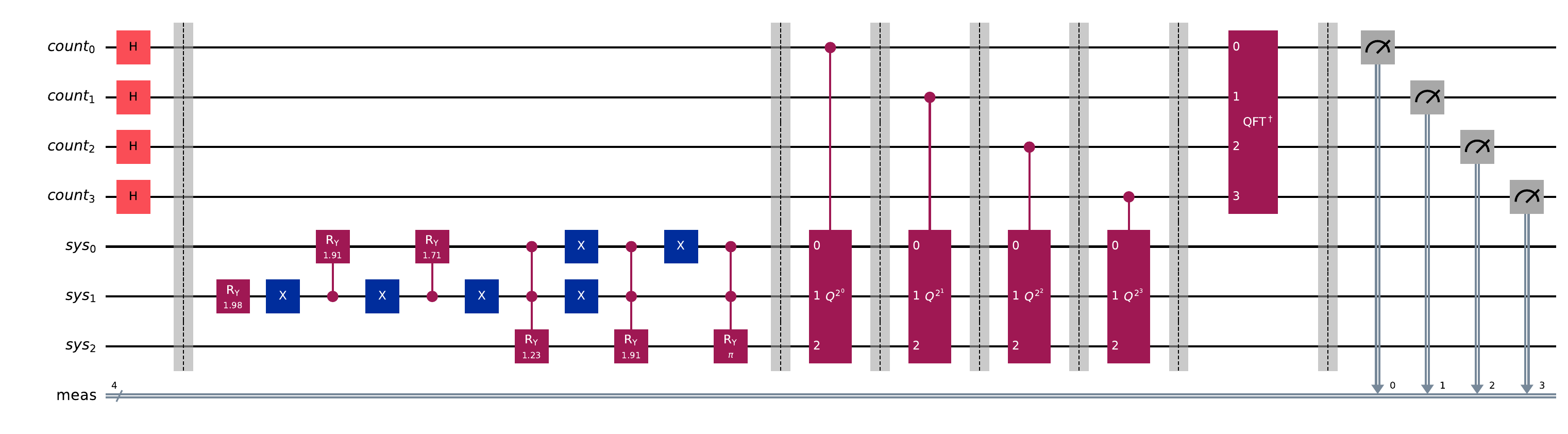}
  \caption{QPE circuit for the $\bar{\nu}$ estimation problem, $m = 4$
           ancilla qubits (QPE register, top) and $n = 3$ system qubits
           (bottom, including ancilla).
           State prep $\mathcal{A}$ loads $P(\nu)$ via an $R_y$ tree and
           $f(\nu)$ via controlled $R_y$ rotations on the ancilla;
           controlled-$\mathcal{Q}^{2^k}$ blocks amplify the good
           amplitude; QFT$^\dagger$ maps the phase to the register outcome.}
  \label{fig:circuit}
\end{figure*}

\paragraph{Circuit.}
Fig.~\ref{fig:circuit} shows the QPE circuit at $m = 4$.
The state-preparation block $\mathcal{A}$ is built in two stages.

\emph{Stage 1 — loading $P(\nu)$.}
With two state qubits $q_1$ (MSB) and $q_0$ (LSB), each basis state
$|\nu\rangle$ corresponds to the binary representation of $\nu$.
A binary $R_y$ tree loads the distribution level by level:
first, $R_y(\theta_1)$ on $q_1$ sets the marginal
$P(q_1{=}1) = P(2){+}P(3) = 0.7$, giving
$\theta_1 = 2\arcsin\!\sqrt{0.7} \approx 1.982$\,rad;
then two controlled $R_y$ gates on $q_0$ handle the
conditional probabilities,
$P(q_0{=}1\,|\,q_1{=}0) = P(1)/[P(0){+}P(1)] = 2/3$ and
$P(q_0{=}1\,|\,q_1{=}1) = P(3)/[P(2){+}P(3)] = 4/7$,
with angles $\theta_{01} = 2\arcsin\!\sqrt{2/3} \approx 1.911$\,rad and
$\theta_{11} = 2\arcsin\!\sqrt{4/7} \approx 1.714$\,rad respectively.
After this stage the state is
$\sum_{\nu=0}^{3}\sqrt{P(\nu)}\,|\nu\rangle$.

\emph{Stage 2 — loading $f(\nu)$.}
A set of doubly-controlled $R_y$ gates on the ancilla qubit $q_2$
rotate it by $\alpha_\nu = 2\arcsin\!\sqrt{f(\nu)}$ conditioned on
each basis state $|\nu\rangle$:
$\alpha_0 = 0$, $\alpha_1 = 2\arcsin(1/\sqrt{3}) \approx 1.23$\,rad,
$\alpha_2 = 2\arcsin\!\sqrt{2/3} \approx 1.91$\,rad,
$\alpha_3 = \pi$ (ancilla fully flipped to $|1\rangle$ when $\nu=3$).
%Control-on-zero conditions are handled by $X$ gates bracketing the controlled rotation.
Control-on-$|0\rangle$ conditions (gate fires when the control qubit is $|0\rangle$) are converted to the standard control-on-$|1\rangle$ by flanking the rotation with $X$ gates.
The combined state after $\mathcal{A}$ matches Eq.~\eqref{eq:stateprep}
exactly.

The Grover iterate $\mathcal{Q}$ is built once as an $8\times 8$
unitary; for $m > 4$, the controlled power $\mathcal{Q}^{2^k}$ is
obtained by $k$ successive matrix squarings rather than $2^k$ circuit
repetitions, keeping circuit build time $O(m)$.
The inverse QFT follows standard decomposition~\cite{Nielsen2000}.

\begin{figure}[!tb]
  \centering
  \includegraphics[width=\columnwidth]{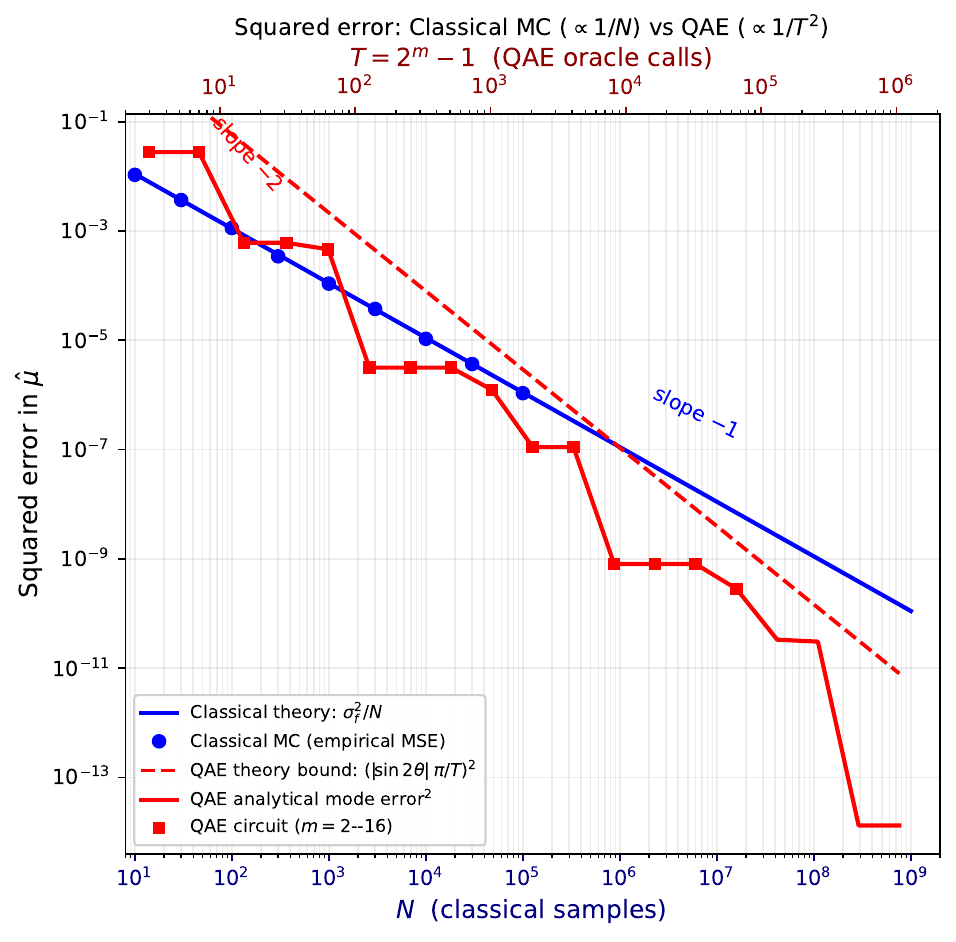}
  \caption{Squared error vs.\ oracle calls $T$ (QAE, red top axis)
           and sample count $N$ (classical MC, blue bottom axis).
           Red squares: Qiskit circuit runs ($m=2$--$16$); for each $m$,
           8192 shots are taken and the mode of $\hat{y}$ is decoded via
           Eq.~\eqref{eq:decode}. Solid red: analytical mode error
           (infinite-shots limit). Dashed red: QAE bound~\eqref{eq:bound}.
           Blue circles: empirical MSE; blue line: $\sigma_f^2/N$.}
  \label{fig:toy}
\end{figure}

\paragraph{Results.}
Fig.~\ref{fig:toy} shows the squared error $|\hat{\mu} - \mu|^2$
versus oracle calls $T$ (QAE, red) and classical sample count $N$ (MC, blue).
The QAE analytical curve is computed from Eq.~\eqref{eq:decode} at each $m$
and matches the decoded mode in these simulations ($m = 2$--$16$, 8192 shots),
consistent with $|\psi\rangle = \mathcal{A}|0\rangle$ lying precisely in the
two-dimensional eigenspace of $\mathcal{Q}$.
The slopes $-2$ (QAE) and $-1$ (classical) on the log-log plot
directly visualize the quadratic advantage.

\subsection{II. U-238 Resonance Integral}
\noindent The resonance integral
\begin{equation}\label{eq:IR}
  I_R = \int_{E_1}^{E_2} \frac{\sigma_a(E)}{E}\,dE
\end{equation}
governs the effective absorption of neutrons slowing down through a
resonance in the $1/E$ flux spectrum.
For U-238, we use the $6.674$\,eV resonance with single-level
Breit-Wigner parameters $E_0 = 6.674$\,eV, $\Gamma = 0.02452$\,eV,
$\sigma_{a,\text{max}} = 21{,}600$\,b~\cite{Mughabghab2006},
and integrate over $[E_0 \pm 10\Gamma]$.

\paragraph{QAE encoding}
We discretize $[E_0 \pm 10\Gamma]$ into $N_\text{bins} = 256$ bins
($n_x = 8$ qubits, $q_0$--$q_7$) with centers
$E_x = E_\text{min} + (x + \tfrac{1}{2})\Delta E$.
%The state-preparation operator $\mathcal{A}$ is built by the same binary-tree technique as the discrete case, now scaled to 8 levels via the 8 qubits, with one additional qubit as ancilla ($q_8$).
The state-preparation operator $\mathcal{A}$ uses the same binary $R_y$ tree: 
each of the $n_x = 8$ qubits $q_0$--$q_7$ adds one level, giving $2^8 = 256$ leaves, one per energy bin.

\emph{Stage 1 — loading $P(x)$.}
The $1/E$ slowing-down spectrum $P(x) \propto 1/E_x$ is encoded
level by level: at level $\ell$ (qubit $q_\ell$), the tree contains
$2^\ell$ nodes, with each holding the conditional probability that the next
bit equals 1 if the $\ell$ bits above it are already set.
The rotation angle at each node is
\begin{equation}\label{eq:tree}
  \theta_\text{node} =
    2\arcsin\!\sqrt{\frac{\sum_{x \in \text{right child}} P(x)}
                        {\sum_{x \in \text{parent}} P(x)}}.
\end{equation}
The full tree requires $\sum_{\ell=0}^{7} 2^\ell = 255$ $R_y$ rotations,
%the deepest of which are controlled on all 7 higher qubits.
the deepest of which are controlled on the 7 preceding qubits $q_0$--$q_6$.

\emph{Stage 2 — loading $f(x)$.}
For each of the 256 bins, the ancilla $q_8$ receives the Breit-Wigner
rotation,
\begin{equation}\label{eq:bwangle}
  \alpha_x = 2\arcsin\!\sqrt{f(x)} =
  2\arcsin\!\frac{\Gamma/2}{\sqrt{(E_x - E_0)^2 + (\Gamma/2)^2}},
\end{equation}
conditioned on $|x\rangle$, requiring 256 8-qubit-controlled $R_y$ gates.

Together, Stages 1 and 2 require $O(2^8)$ multi-controlled gates. 
In practice, Qiskit's \texttt{initialize()} decomposes the 9-qubit state preparation into a large number of primitive 2-qubit gates, 
making gate-level simulation prohibitively slow in this case.
Instead, $|\psi\rangle = \mathcal{A}|0\rangle$ is constructed analytically
as a 512-element vector and the $512\times512$ Grover matrix formed via the
rank-1 identity:
 since $\mathcal{A}\,S_0\,\mathcal{A}^\dagger = I - 2|\psi\rangle\langle\psi|$,
$\mathcal{Q} = \mathrm{diag}(\mathbf{s}) - 2|\psi\rangle\langle\psi|\mathbf{s}$,
where $s_i = +1$ (ancilla~$= 0$) or $-1$ (ancilla~$= 1$).
QPE is simulated by eigendecomposing $\mathcal{Q}$ once, then evaluating
the measurement distribution via a batched matrix multiply and Fast Fourier Transformation (FFT). 
This is mathematically identical to the circuit, but bypassing gate decomposition.
Then $\mu = \mathbb{E}[f(X)]$ and
$I_R = \mu \cdot C_\text{phys}$,
where $C_\text{phys} = \sigma_{a,\text{max}} \,\Delta E \sum_x 1/E_x \approx 1578$\,b
is a classically precomputed constant.

\begin{figure}[!tb]
  \centering
  \includegraphics[width=\columnwidth]{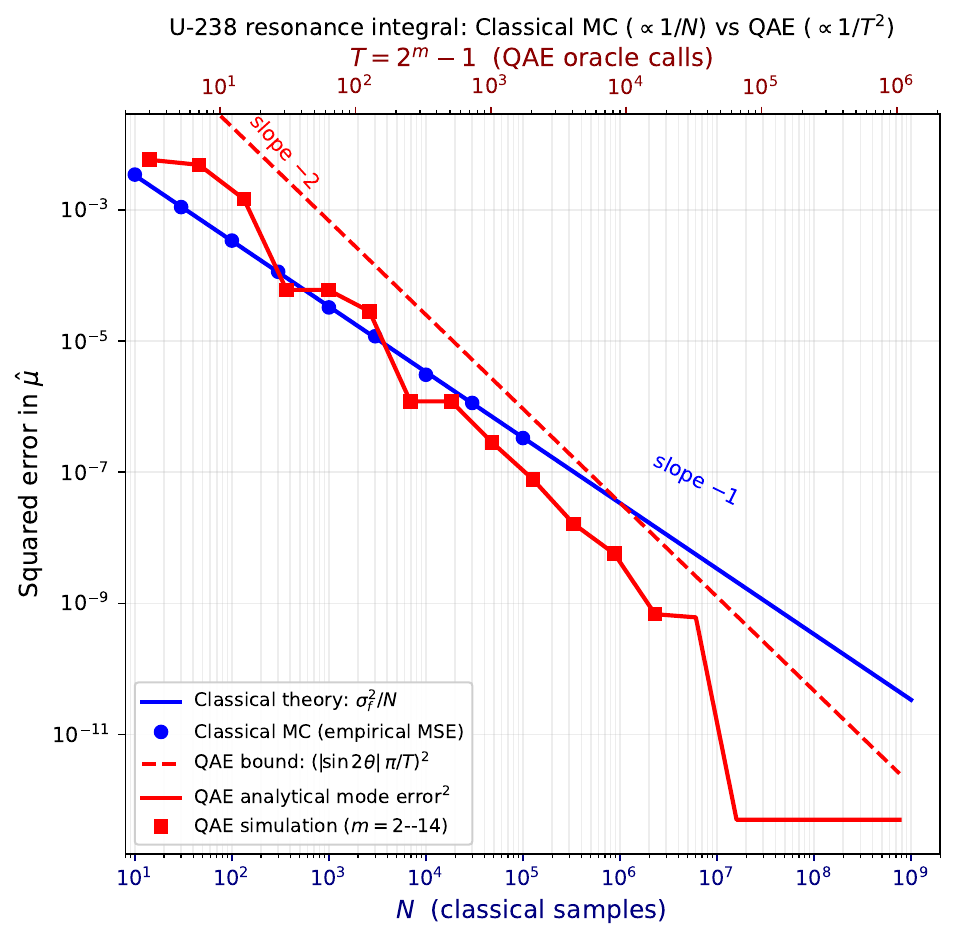}
  \caption{Squared error in $\hat{\mu}$ for the U-238 resonance integral.
           Axes and line styles as in Fig.~\ref{fig:toy}; red squares
           use the eigendecomposition simulation with 8192 shots (mode
           of $\hat{y}$) per $m$.
           At $m = 14$ ($T=16{,}383$), $\hat{I}_R = 120.65$\,b
           vs.\ true $120.69$\,b ($0.03\%$ error).}
  \label{fig:resonance}
\end{figure}

\paragraph{Results.}
The true value is $\mu_\text{true} = 0.07648$,
$I_{R,\text{true}} = 120.69$\,b.
Fig.~\ref{fig:resonance} shows the MSE in $\hat{\mu}$ versus
$T = 2^m - 1$ oracle calls (QAE) and versus classical sample count $N$
(MC sampling from $P(x)$).
The quadratic convergence advantage is again clearly visible.
At $m = 14$ ($T = 16{,}383$ oracle calls) the estimate
$\hat{\mu} = 0.07645$ gives $I_R = 120.65$\,b, a $0.03\%$ error.
The MSE in $\hat{\mu}$ at $m=14$ is $6.86\times10^{-10}$, while
matching this with classical MC requires 
$N = \sigma_f^2/\mathrm{MSE} \approx 49{,}400{,}000$
samples, 
giving a speedup ratio of $N/T \approx 3{,}016$, 
over three orders of magnitude.

%%%%%%%%%%%%%%%%%%%%%%%%%%%%%%%%%%%%%%%%%%%%%%%%%%%%%%%%%%%%%%%%%%%%%
\section{Conclusions}

QPE-based QAE has been implemented on a Qiskit~2.2 statevector
simulator and applied to two nuclear-relevant expectation-value problems.
In both cases, the MSE falls with slope $-2$ on a log-log
plot versus oracle calls, confirming the $O(1/T^2)$ convergence
predicted by theory and matching the analytical mode-error formula
in both problems.
Notably, for the U-238 resonance integral, QAE recovers $I_R = 120.65$\,b at
$0.03\%$ accuracy with $m = 14$, using an order-of-magnitude fewer
equivalent function evaluations than classical MC at the same precision.
A key implementation limitation is that the gate-level circuit becomes
infeasible for $n \geq 9$ system qubits on current quantum simulators due to the
exponential cost of state-preparation decomposition. However, the eigendecomposition
surrogate method used here is an accurate alternative for simulation but does not
produce a hardware-ready circuit.
Future work will implement fault-tolerant quantum arithmetic circuits for
the Breit-Wigner state preparation~\cite{Stamatopoulos2020}, enabling
scaling to $n \geq 18$ bins and multiple resonances, and will explore
iterative QAE variants~\cite{Montanaro2015} to account for near-term
noisy hardware.

%%%%%%%%%%%%%%%%%%%%%%%%%%%%%%%%%%%%%%%%%%%%%%%%%%%%%%%%%%%%%%%%%%%%%
\section*{Acknowledgments}

The authors thank the Penn State Institute for Computational and Data Sciences
(ICDS) for providing the high-performance computing resources used in this work.
M.\ Jin acknowledges support from the ICDS Faculty Upskilling Fellowship.

%%%%%%%%%%%%%%%%%%%%%%%%%%%%%%%%%%%%%%%%%%%%%%%%%%%%%%%%%%%%%%%%%%%%%
\bibliographystyle{ans}
\bibliography{qmc_ans}

@article{Brassard2002,
  author  = {Brassard, Gilles and H{\o}yer, Peter and Mosca, Michele and Tapp, Alain},
  title   = {Quantum Amplitude Amplification and Estimation},
  journal = {Contemporary Mathematics},
  volume  = {305},
  pages   = {53--74},
  year    = {2002},
  doi     = {10.1090/conm/305/05215}
}

@article{Montanaro2015,
  author  = {Montanaro, Ashley},
  title   = {Quantum speedup of {M}onte {C}arlo methods},
  journal = {Proceedings of the Royal Society A},
  volume  = {471},
  number  = {2181},
  pages   = {20150301},
  year    = {2015},
  doi     = {10.1098/rspa.2015.0301}
}

@article{Woerner2019,
  author  = {Woerner, Stefan and Egger, Daniel J.},
  title   = {Quantum risk analysis},
  journal = {npj Quantum Information},
  volume  = {5},
  pages   = {15},
  year    = {2019},
  doi     = {10.1038/s41534-019-0130-6}
}

@article{Stamatopoulos2020,
  author  = {Stamatopoulos, Nikitas and Egger, Daniel J. and Sun, Yue and
             Zoufal, Christa and Iten, Raban and Shen, Ning and Woerner, Stefan},
  title   = {Option Pricing using Quantum Computers},
  journal = {Quantum},
  volume  = {4},
  pages   = {291},
  year    = {2020},
  doi     = {10.22331/q-2020-07-06-291}
}

@article{Martin2012,
  author  = {Martin, William R.},
  title   = {Challenges and Prospects for Whole-Core {M}onte {C}arlo Analysis},
  journal = {Nuclear Engineering and Technology},
  volume  = {44},
  number  = {2},
  pages   = {151--160},
  year    = {2012},
  doi     = {10.5516/NET.01.2012.502}
}

@book{Nielsen2000,
  author    = {Nielsen, Michael A. and Chuang, Isaac L.},
  title     = {Quantum Computation and Quantum Information},
  publisher = {Cambridge University Press},
  year      = {2000}
}

@article{Qiskit2024,
  author  = {Javadi-Abhari, Ali and Treinish, Matthew and Krsulich, Kevin and
             Wood, Christopher J. and Lishman, Jake and Gacon, Julien and
             Martiel, Simon and Nation, Paul D. and Bishop, Lev S. and
             Cross, Andrew W. and Johnson, Blake R. and Gambetta, Jay M.},
  title   = {Quantum computing with {Qiskit}},
  journal = {arXiv preprint},
  year    = {2024},
  doi     = {10.48550/arXiv.2405.08810}
}

@article{Olivier2024,
  author  = {Olivier, No\'{e} and Nowak, Michel},
  title   = {Monte {C}arlo particle transport on quantum computers},
  journal = {arXiv preprint},
  year    = {2024},
  doi     = {10.48550/arXiv.2410.19489}
}

@book{Mughabghab2006,
  author    = {Mughabghab, S. F.},
  title     = {Atlas of Neutron Resonances: Resonance Parameters and Thermal
               Cross Sections. {Z}=1--100},
  publisher = {Elsevier},
  edition   = {5},
  year      = {2006}
}

\end{document}